\documentclass[twocolumn,aps,pra,showpacs
]{revtex4-2}
\usepackage{eurosym}
\usepackage{epstopdf}
\usepackage{latexsym}
\usepackage[colorlinks=true,citecolor=blue,linkcolor=magenta]{hyperref}
\usepackage{graphicx}
\usepackage{color, soul}
\usepackage{amsmath}
\usepackage{natbib}
\usepackage{bm}
\usepackage{times}
\usepackage{url}
\usepackage{amssymb}
\usepackage{braket}
\usepackage{mathtools}
\usepackage{commath}
\usepackage{bigints}
\usepackage{booktabs}
\usepackage{multirow}
\usepackage{epstopdf}
\usepackage{pdfcomment}
\usepackage[normalem]{ulem}
\usepackage{mathrsfs}
\usepackage{xr}
\usepackage{cleveref}
\usepackage[top=30mm, bottom=25mm, left=25mm, right=30mm]{geometry}
\usepackage{babel}
\usepackage[utf8]{inputenc}
\usepackage[mathscr]{euscript}
\usepackage{silence}

\graphicspath{{images/}}

\inputencoding{utf8}
\definecolor{dgreen}{rgb}{0.0, 0.5, 0.0}
\def\be{\begin{equation}}
\def\ee{\end{equation}}
\def\bea{\begin{eqnarray}}
\def\eea{\end{eqnarray}}

\def\bi{\begin{itemize}}
\def\ei{\end{itemize}}
\def\ben{\begin{enumerate}}
\def\een{\end{enumerate}}

\begin{document}

\title{Temporal localization of optical waves supported by a copropagating
quasiperiodic structure}
\author{Majid Yazdani-Kachoei$^{1}$}
\author{Krzysztof Sacha$^{1,2}$}
\author{Boris A. Malomed$^{3,4}$}
\affiliation{$^1$Instytut Fizyki Teoretycznej, Wydzial Fizyki, Astronomii i
Informatyki Stosowanej, Uniwersytet Jagiello\'nski, ulica Profesora
Stanislawa Lojasiewicza 11, PL-30-348 Krak\'{o}w, Poland}
\affiliation{$^2$Centrum Marka Kaca, Uniwersytet Jagiello\'nski, ulica Profesora
Stanisława Łojasiewicza 11, PL-30-348 Kraków, Poland}
\affiliation{$^3$Department of Physical Electronics, School of Electrical
Engineering, Faculty of Engineering, Tel Aviv University, Tel Aviv 69978,
Israel} 
\affiliation{$^4$Instituto de Alta Investigaci\'{o}n, Universidad de
Tarapac\'{a}, Casilla 7D, Arica, Chile}

\begin{abstract}
Research on time crystals concerns the spontaneous breaking of translational
symmetry in time, as well as the realization of phenomena and phases known
from solid-state physics in the time domain. Periodically driven systems of
massive particles are widely used in these studies. In the present work, we
consider a photonic system and demonstrate that stable nonlinear propagation
of a strong optical wave in a fiber leads to the establishment of
quasi-periodic oscillations in the electromagnetic field intensity. A
second, weaker signal optical wave propagating in the fiber senses these
oscillations and, as a result, undergoes exponential localization in time.
This is a temporal analogue of Aubry-Andr\'{e} localization. If an optical
detector is placed at a certain position in the fiber, the temporal
localization of the probe wave will be observed in the form of the signal
which emerges and then decays as a function of time.
\end{abstract}

\date{\today }
\maketitle


\section{Introduction}

A quantum particle in the presence of a stationary potential with a
disordered spatial structure may exhibit Anderson localization (AL), which
gives rise to eigenstates exponentially localized at different positions in
space \cite{Anderson1958,Flach}. AL was studied in a broad range of
phenomenology modeled by disordered wave equations, ranging from the sound
propagation to matter waves \cite{Lagendijk2009}. In particular, this effect
was predicted and observed in the propagation of light \cite{Wiersma,Segev}
and plasmonic \cite{Stockman} excitations in random optical and photonic
media \cite{Segev2,Mafi,Yamilov}. Typically, AL is observed in the
configuration space, resulting from the presence of time-independent spatial
disorder. It can also be observed in the momentum space of a quantum
particle. In that case, it is associated with the quantum suppression of
classical diffusion in systems that are classically chaotic \cite%
{Fishman:LocDynAnders:PRL82,Lemarie:Anderson3D:PRA09}.

Recently, it has been shown that AL is also possible in the time domain
\cite{Sacha15a,sacha16,delande17,SachaTC2020}. This phenomenon belongs to
the field of time crystals, which has been intensively elaborated over the
past decade \cite{Sacha2017rev,SachaTC2020,GuoBook2021,Zaletel2023}. Akin to
their spatial counterparts, {time crystals }exhibit spontaneous regular
self-organization -- not in space, but in the time domain \cite%
{Wilczek2012,Bruno2013b,Watanabe2015,Sacha2015,Khemani16,ElseFTC,Zhang2017,
Choi2017,Pal2018,Rovny2018,Smits2018,Autti2018,Mi2022,Randall2021,Frey2022,Kessler2020,Kyprianidis2021,Xu2021,Autti2021,Taheri2020,Taheri2022a,Taheri2022b,Liu2023,Bao2024,Kazuya2024,Liu2024,Liu2024a,kopaei2023}
. This self-organization involves spontaneous breaking of the
time-translation symmetry, thus being a temporal counterpart of one of the
most important features of solid-state physics. In {time-crystal} setups it
is also possible to realize a wide range of phases and phenomena known in
condensed matter physics, which are not necessarily associated with the
spontaneous breaking of the time-translation symmetry \cite%
{SachaTC2020,GuoBook2021,Hannaford2022,Giergiel2024,Kopaei2024a}. These
include, in particular, AL in the time domain \cite%
{Sacha15a,sacha16,delande17,SachaTC2020}: a quantum particle perturbed by a
fluctuating external force reveals localization in time, meaning that, if
one places a detector at a certain spatial position, temporal variations of
the detection probability exhibit exponential localization around a certain
moment in time. The description of this phenomenon reduces to the Anderson
model.

Temporal disorder has also been investigated in photonic time crystals.
These are spatially homogeneous dielectric materials where the refractive
index is periodically modulated in time. Solutions of the respective Maxwell
equations reveal band structures in the momentum space \cite%
{Biancalana2007,Zurita2009,Galiffi2022}.
Essentially, the temporal disorder leads to an unbounded amplification of
electromagnetic pulses over time and slowdown of their propagation through
the material \cite{Pendry1982,Sharabi2021,Carminati2021,Apffel2021,Zhou2024}.

In the three-dimensional (3D) space, AL is present if the underlying
disorder is sufficiently strong -- namely, 3D eigenstates with a given
energy localize if the disorder strength exceeds a critical value \cite%
{Abrahams:Scaling:PRL79}. In 2D and 1D cases, even arbitrarily weak disorder
leads to AL, although in 2D the localization length of the wave function of
the quantum particle may be very large. On the other hand, in the 1D case,
replacing the random spatial potential by a lattice (periodic) one, which is
subject to superlattice modulation with an incommensurate spatial period,
one arrives at the Aubry-Andr\'{e} (AA) model with an effective \textit{%
quasiperiodic} (QP)\ potential \cite{Aubry1980}. In spite of the absence of
disorder in the AA model, sufficiently strong modulation also leads to the
localization of eigenstates of the quantum particle.

In its strict sense, the AA model is a tight-binding discrete one
characterized by rate of tunneling between sites of a 1D chain and on-site
energies that vary in a QP manner along the chain. For fixed on-site
energies, the localization of eigenstates takes place if the tunneling rate
falls below a certain critical value, i.e., the delocalizing overlap between
adjacent sites is not too strong \cite{Aubry1980}. While our system is
continuous, rather than discrete one, we demonstrate that the localization
mechanism acting in the system is quite similar to its counterpart in the AA
model.

Fiber optics offers a specific possibility to create a specific setting for
the localization in the temporal domain, instead of the spatial one. Namely,
one can consider copropagation of two waves carried by two different
channels in the same fiber, which represent different wavelengths. In the
\textit{support channel}, the wave is assumed to be strong, hence it
propagates in the nonlinear regime, with the corresponding carrier
wavelength belonging to the range of normal group-velocity dispersion (GVD)
\cite{Agrawal}. Then, this wave can develop a stable periodically modulated
shape. Roughly speaking, it is a chain of dark solitons, alias a\textit{\
cnoidal} wave, so called because it is represented by an exact solution of
the corresponding nonlinear Schr\"{o}dinger (NLS) equation expressed in
terms of Jacobi elliptic functions, such as $\mathrm{cn}$ \cite{cnoidal}.
Through the cross-phase-modulation (XPM) effect, the cnoidal wave induces an
effective periodic potential in the second, \textit{signal channel}, which
carries optical signals. In particular, in Refs.~\cite{Ship1,Ship2} the
situation was considered when the signal channel operated with anomalous
GVD, thus carrying signals in the form of bright solitons (a similar, but
less stable, scheme with a chain of bright solitons created in the support
channel with anomalous GVD was recently considered in Ref.~\cite{Agr-OL}). A
well-known problem in fiber-optic telecommunications is the temporal jitter,
alias the Gordon-Haus effect \cite{GH}, i.e., random walk of the solitons in
the temporal domain, caused by their interaction with random optical noise
spontaneously emitted by amplifiers, which must be periodically inserted in
the long fiber link. Although the jitter does not destroy solitons, it tends
to obliterate data encoded by temporal positions of the solitons in the
signal stream. As proposed in Refs.~\cite{Ship1,Ship2}, the effective
XPM-induced periodic (cnoidal-like) potential, acting on the solitons in the
signal channel, may effectively suppress the jitter, thus stabilizing the
data transmission by the soliton stream.

The two-channel support-signal scheme, that can be implemented in the
optical fiber, suggests a possibility to induce an effective QP (rather than
periodic) potential in the signal channel. To this end, the carrier
wavelength in the support channel should be taken close enough to the
zero-dispersion point, which makes it necessary to take into regard the
third-order GVD, in addition to the usual second-order normal-GVD term \cite%
{Agrawal}. Indeed, the NLS equation which includes solely the second-order
GVD is commonly known to be integrable \cite{ZS}. This fact make the
effective second-order ordinary differential equation (ODE) for the wave
amplitude in the support channel also integrable. As a result, the
integrable ODE, being formally equivalent to the mechanical equation of
motion for a particle in an external potential well, is also integrable,
producing only periodic solutions of the cnoidal type (the soliton is a
limit case of the solution with an infinite period). On the other hand, the
addition of the terms accounting for the third-order GVD breaks the
integrability of the NLS equation, and makes its ODE reduction for
stationary waves also non-integrable. As we demonstrate below, the
stationary waves exist, in the support channel, as robust solutions with a
QP structure. Then, by dint of the XPM effect, they induce an effective QP
potential in the signal channel. We consider the wave in the signal channel
as a weak one, governed by the linear propagation equation. Under the action
of the XPM-induced QP potential, the signal wave develops the
temporal-domain localization, similar to that in the spatial-domain AA model.

The following presentation is organized as follows: the two-channel model is
introduced in Section~\ref{sec2}, where we also formulate the framework for
the analysis of the QP solutions in terms of the single NLS equation, which
represents the support channel. The equation includes the second- and
third-order GVD terms (with the positive second-order GVD coefficient, which
corresponds to the normal GVD, as, in the case of the anomalous GVD, the QP
waves may be subject to the modulational instability) and the cubic Kerr
nonlinearity. The complex ODE which produces QP waves in the support channel
is tantamount to a real dynamical system with a six-dimensional phase space.
We derive two dynamical invariants of the six-dimensional system, \textit{viz%
}., its Hamiltonian and an invariant corresponding to the phase invariance
of the complex ODE. The situation seems as the one for a system with three
degrees of freedom and two dynamical invariants, which implies that the
system is non-integrable. Therefore, it is able to produce QP solutions. In
Section~\ref{sec3} numerical solutions are produced: first, QP waves in the
support channel, then solutions of the linear Schr\"{o}dinger equation in
the signal channel, with the effective XPM-induced QP potential. The latter
solutions demonstrate the localization of the signal wave function in the
time domain, which is main result reported by the present work.
The paper is concluded by Section~\ref{sec4}.

\section{The model and support structure: analytical considerations}

\label{sec2}

The system of NLS equations for the copropagation of optical waves with
envelope amplitudes $U(z,t)$ and $\Psi (z,t)$ in the support and signal
channels, respectively, carried by different wavelengths in the fiber, is
written as \cite{Agrawal,Ship1,Ship2}

\begin{gather}
i\frac{\partial U}{\partial z}=\left[ ic\frac{\partial }{\partial t}+\frac{%
\beta _{2}}{2}\frac{\partial ^{2}}{\partial t^{2}}+\frac{i\beta _{3}}{6}%
\frac{\partial ^{3}}{\partial t^{3}}-|U(z,t)|^{2}\right] U,  \label{U} \\
i\frac{\partial \Psi }{\partial z}=\left[ \frac{\gamma _{2}}{2}\frac{%
\partial ^{2}}{\partial t^{2}}+\frac{i\gamma _{3}}{6}\frac{\partial ^{3}}{%
\partial t^{3}}-2|U(z,t)|^{2}\right] \Psi ,  \label{Psi}
\end{gather}%
where $z$ is the propagation distance, $t\equiv \tau -z/V_{\mathrm{gr}}$ is
the reduced (local) time, defined for the signal channel which is carried by
the electromagnetic wave with group velocity $V_{\mathrm{gr}}$ ($\tau $ is
time per se), real $c$ is the group-velocity mismatch between the channels,
while $\beta _{2,3}$ and $\gamma _{2,3}$ are real coefficients of the
second- and third-order GVD in the support and signal channels,
respectively. Recall that the cases of $\beta _{2},\gamma _{2}>0$ and $<0$
are classified, respectively, as normal or anomalous second-order GVD \cite%
{Agrawal}. The cubic term in Eq.~(\ref{U}) represents the usual SPM
(self-phase-modulation) nonlinearity acting in the support channel, and the last term in Eq.~(\ref{Psi}) accounts for the action of XPM in the signal channel.

In Refs.~\cite{Ship1,Ship2}, the carrier wavelengths of the support and
signal channels were chosen symmetrically with respect to the
zero-dispersion point, which implies $\beta _{2}=-\gamma _{2}$ and $c=0$. In
the case of small $c\neq 0$, the respective term can be eliminated from Eq.~(%
\ref{U}) by means of the substitution which does not affect Eq.~(\ref{Psi}),
\textit{viz}., $U(z,t)\equiv \tilde{U}(z,t)\exp \left[ -\left( c^{2}/\beta
_{2}\right) z-i\left( c/\beta _{2}\right) t\right] $, $\beta _{2}\rightarrow
\beta _{2}+\left( \beta _{3}/\beta _{2}\right) c$. Therefore, we drop the
term $ic\partial U/\partial t$ in Eq.~(\ref{U}).

As said above, the nonlinear (XPM and SPM) effects produced by the weak wave
on the support and signal channels are neglected in Eqs.~(\ref{U}) and (\ref%
{Psi}). The coefficients accounting for the SPM and XPM effects (with the
usual ratio, $\mathrm{XPM/SPM}=2$ \cite{Agrawal}) of the strong wave on the
support and signal channels are set to be $1$ and $2$, respectively, by
means of rescaling.

Because Eq.~(\ref{U}) does not couple to Eq.~(\ref{Psi}), the former
equation can be solved separately. Once the solution is known, we proceed to
solving the latter equation, where $|U(z,t)|^{2}$ acts as an external
potential. Aiming to realize the AA localization in the temporal domain, we
are interested in a QP effective potential. To investigate this possibility,
stationary solutions to Eq.~(\ref{U}) with an arbitrary real propagation
constant $k$ are looked for as
\begin{equation}
U\left( z,t\right) =e^{ikz}u(t),  \label{UV}
\end{equation}%
where complex function $u(t)$ satisfies the third-order ODE:
\begin{equation}
ku+\frac{1}{2}\beta _{2}\frac{d^{2}u}{dt^{2}}+\frac{i\beta _{3}}{6}\frac{%
d^{3}u}{dt^{3}}-|u|^{2}u=0,  \label{ODE}
\end{equation}%
which is tantamount to a real dynamical system in the six-dimensional phase
space, if the complex variable $u(t)$ is split into the real and imaginary
parts, $a_{1}$ and $a_{2}$:
\begin{equation}
u(t)\equiv a_{1}(t)+ia_{2}(t).  \label{i}
\end{equation}%
To conclude what kind of solutions may be expected (periodic, quasiperiodic,
random, {or unbounded}), it is crucially important to identify dynamical
invariants (conserved quantities) of Eq.~(\ref{ODE}).

First, to identify the dynamical invariant, which is related, by way of the
Noether theorem \cite{Mechanics}, to the invariance of Eq.~(\ref{ODE}) with
respect to an arbitrary phase shift of the complex function $u(t)$, we note
that Eq.~(\ref{ODE}) can be derived from the corresponding real Lagrangian:
\begin{gather}
\mathcal{L}=k|u|^{2}-\frac{1}{2}\beta _{2}\left\vert \frac{du}{dt}%
\right\vert ^{2}  \notag \\
+\frac{i\beta _{3}}{12}\left[ \left( \frac{d^{2}u}{dt^{2}}\right) ^{\ast }%
\frac{du}{dt}-\frac{d^{2}u}{dt^{2}}\left( \frac{du}{dt}\right) ^{\ast }%
\right] -\frac{1}{2}|u|^{4},  \label{Lagr}
\end{gather}%
where $\ast $ stands for the complex conjugate. Further, the Lagrangian can
be written in terms of the Madelung substitution,
\begin{equation}
u(t)=A(t)\exp \left( i\phi (t)\right) ,  \label{Madelung}
\end{equation}%
with real amplitude $A(t)\equiv \left\vert u(t)\right\vert $ and phase $\phi
(t)$. The result is
\begin{gather}
\mathcal{L}=kA^{2}-\frac{1}{2}A^{4}-\frac{1}{2}\beta _{2}\left[ \left( \frac{%
dA}{dt}\right) ^{2}+A^{2}\left( \frac{d\phi }{dt}\right) ^{2}\right]  \notag
\\
+\frac{\beta _{3}}{6}\left[ 2\left( \frac{dA}{dt}\right) ^{2}\frac{d\phi }{dt%
}-A\frac{d^{2}A}{dt^{2}}\frac{d\phi }{dt}\right.  \notag \\
\left. +A^{2}\left( \frac{d\phi }{dt}\right) ^{3}+A\frac{dA}{dt}\frac{%
d^{2}\phi }{dt^{2}}\right] .  \label{L2}
\end{gather}%
Lagrangian (\ref{L2}) can be further transformed by replacing the last term
by the one produced by the integration by parts, if one considers the
corresponding integral for the action, $\int \mathcal{L}dt$, \textit{viz}.,
\begin{equation}
A\frac{dA}{dt}\frac{d^{2}\phi }{dt^{2}}\rightarrow -\frac{d}{dt}\left( A%
\frac{dA}{dt}\right) \frac{d\phi }{dt},
\end{equation}%
where
\begin{equation}
\frac{d}{dt}\left( A\frac{dA}{dt}\right) \frac{d\phi }{dt}\equiv \left[
\left( \frac{dA}{dt}\right) ^{2}+A\frac{d^{2}A}{dt^{2}}\right] \frac{d\phi }{%
dt}.
\end{equation}%
The respectively transformed Lagrangian is
\begin{eqnarray}
\mathcal{\bar{L}}=kA^{2}-\frac{1}{2}A^{4}-\frac{1}{2}\beta _{2}\left[ \left(
\frac{dA}{dt}\right) ^{2}+A^{2}\left( \frac{d\phi }{dt}\right) ^{2}\right] &&
\notag \\
+\frac{\beta _{3}}{6}\left[ \left( \frac{dA}{dt}\right) ^{2}\frac{d\phi }{dt}%
-2A\frac{d^{2}A}{dt^{2}}\frac{d\phi }{dt}+A^{2}\left( \frac{d\phi }{dt}%
\right) ^{3}\right]. && \cr &&  \label{L3}
\end{eqnarray}%
Finally, the standard variational procedure, applied to Lagrangian (\ref{L3}%
), demonstrates that the dynamical invariant sought for is
\begin{gather}
I\equiv \frac{\partial \mathcal{\bar{L}}}{\partial \left( d\phi /dt\right) }%
=-\beta _{2}A^{2}\frac{d\phi }{dt}  \notag \\
+\frac{\beta _{3}}{6}\left[ \left( \frac{dA}{dt}\right) ^{2}-2A\frac{d^{2}A}{%
dt^{2}}+3A^{2}\left( \frac{d\phi }{dt}\right) ^{2}\right] .  \label{I}
\end{gather}%
In the case of $\beta _{3}=0$ this dynamical invariant is a commonly known
one (the angular momentum in the plane of coordinates $\left(
a_{1},a_{2}\right) $, see Eq. (\ref{i})), while, to the best of our
knowledge, it was not previously reported in the case of $\beta _{3}\neq 0$.

The second dynamical invariant of Eq.~(\ref{ODE}) is its Hamiltonian. To
derive it, one can use the representation of the solution in the form of
Eq.~(\ref{i}), instead of the Madelung form (\ref{Madelung}). The
substitution of this in the underlying expression (\ref{Lagr}) yields the
Lagrangian in another form:
\begin{gather}
\mathcal{L}=k\left( a_{1}^{2}+a_{2}^{2}\right) -\frac{1}{2}\left(
a_{1}^{2}+a_{2}^{2}\right) ^{2}  \notag \\
-\frac{1}{2}\beta _{2}\left[ \left( \frac{da_{1}}{dt}\right) ^{2}+\left(
\frac{da_{2}}{dt}\right) ^{2}\right]  \notag \\
+\frac{\beta _{3}}{6}\left( \frac{da_{1}}{dt}\frac{d^{2}a_{2}}{dt^{2}}-\frac{%
da_{2}}{dt}\frac{d^{2}a_{1}}{dt^{2}}\right) .  \label{L4}
\end{gather}%
To derive the Hamiltonian from Lagrangian (\ref{L4}), it is necessary to
define momentum-like variable,
\begin{equation}
b_{1,2}\equiv \frac{da_{1,2}}{dt},  \label{ba}
\end{equation}%
in addition to coordinates $a_{1,2}$. Then, it is easy to check that the
correct equations for $a_{1,2}$, together with relation (\ref{ba}), can be
derived from the following modification of the Lagrangian, written in terms
of both $a_{1,2}$ and $b_{1,2}$:
\begin{gather}
\mathcal{L}_{ab}=k\left( a_{1}^{2}+a_{2}^{2}\right) -\frac{1}{2}\left(
a_{1}^{2}+a_{2}^{2}\right) ^{2}+\frac{1}{2}\beta _{2}\left(
b_{1}^{2}+b_{2}^{2}\right)  \notag \\
-\beta _{2}\left( \frac{da_{1}}{dt}b_{1}+\frac{da_{2}}{dt}b_{2}\right) -%
\frac{\beta _{3}}{12}\left( b_{1}\frac{db_{2}}{dt}-b_{2}\frac{db_{1}}{dt}%
\right)  \notag \\
+\frac{\beta _{3}}{6}\left( \frac{da_{1}}{dt}\frac{db_{2}}{dt}-\frac{da_{2}}{%
dt}\frac{db_{1}}{dt}\right) .  \label{L5}
\end{gather}%
Finally, the canonical form of Lagrangian (\ref{L5}), which includes only
the first derivatives, makes it possible to construct the conserved
Hamiltonian by means of the canonical Legendre transformation \cite%
{Mechanics}:
\begin{eqnarray}
H &=&\sum_{j=1}^{2}\left[ \frac{\partial \mathcal{L}_{ab}}{\partial \left(
da_{j}/dt\right) }\frac{da_{j}}{dt}+\frac{\partial \mathcal{L}_{ab}}{%
\partial \left( db_{j}/dt\right) }\frac{db_{j}}{dt}\right] -\mathcal{L}_{ab} %
\cr &=& -k\left( a_{1}^{2}+a_{2}^{2}\right) +\frac{1}{2}\left(
a_{1}^{2}+a_{2}^{2}\right) ^{2}-\frac{1}{2}\beta _{2}\left(
b_{1}^{2}+b_{2}^{2}\right) \cr && +\frac{\beta _{3}}{6}\left( \frac{da_{1}}{%
dt}\frac{db_{2}}{dt}-\frac{da_{2}}{ dt}\frac{db_{1}}{dt}\right) .  \label{H}
\end{eqnarray}
This expression for the Hamiltonian of Eq.~(\ref{ODE}) with $\beta _{3}\neq
0 $ has not been reported previously either, to the best of our knowledge.

Thus, the dynamical system in the six-dimensional phase space, which
corresponds to Eq. (\ref{ODE}), maintains two dynamical invariants, given by
expressions (\ref{I}) and (\ref{H}). For this reason, as mentioned above,
the system \emph{is not integrable} (on the contrary to its commonly known
four-dimensional limit with $\beta _{3}=0$), hence its bounded generic
solutions should be quasiperiodic and/or chaotic.

\section{Numerical results}

\label{sec3}


\subsection{Quasi-periodic waves in the support channel}

\begin{figure*}[t]
\includegraphics[width=\linewidth]{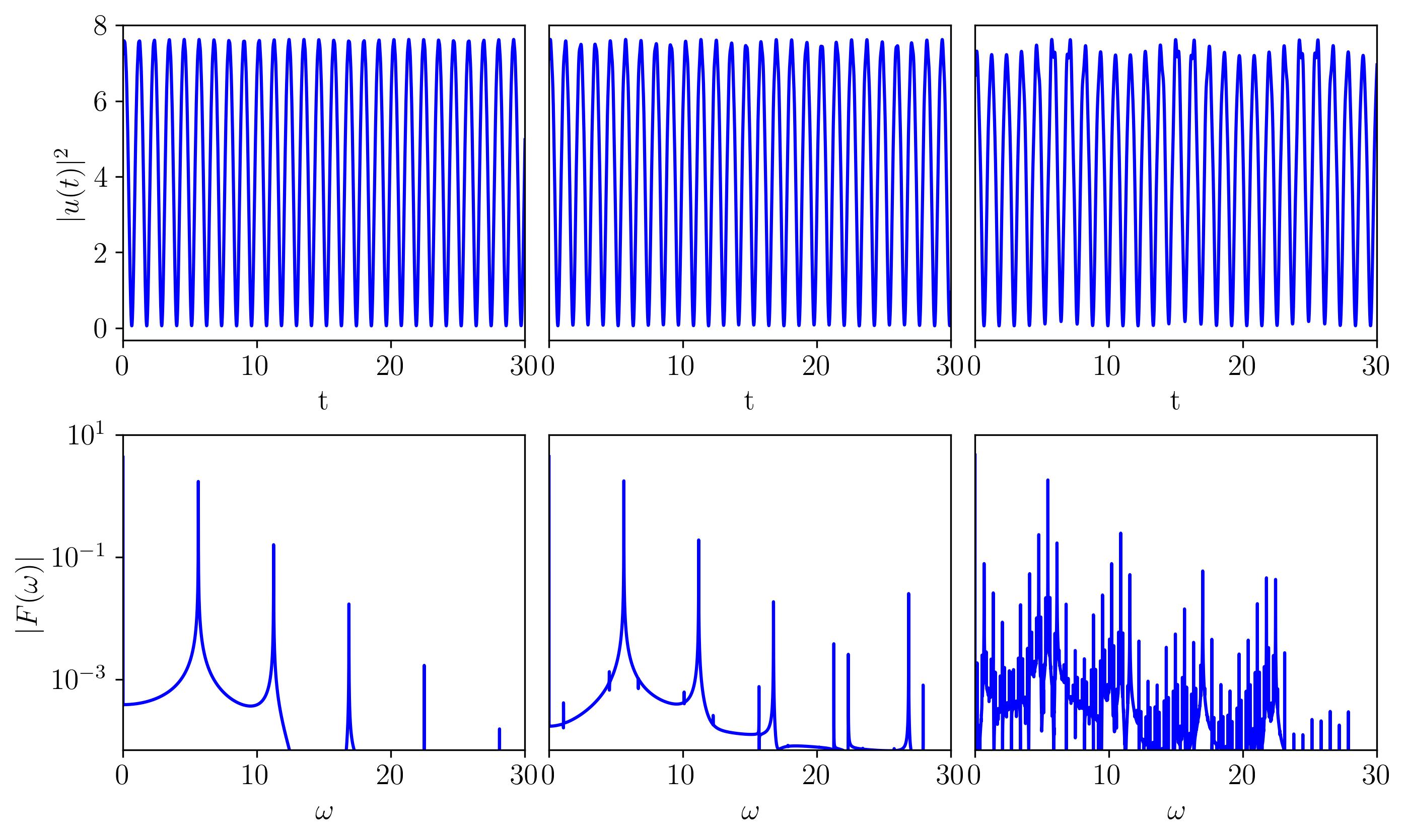}
\caption{Top panels show the effective supporting potentials, $|u(t)|^{2}$,
produced by the numerical solution of Eq.~(\protect\ref{ODE}) with initial
conditions $u(t=0)=2.5+i$, $du/dt(t=0)=d^{2}u/dt^{2}(t=0)=1+i$, and
parameters $\protect\beta _{2}=1$, $k=10$ . The third-order GVD coefficient
takes values $\protect\beta _{3}=0.05$, $\protect\beta _{3}=0.1$, and $%
\protect\beta _{3}=0.15$ in the left, central, and right columns,
respectively. Bottom panels show the respective spectra $|F(\protect\omega %
)| $ on the logarithmic scale, see Eqs. (\protect\ref{Fourier}).}
\label{fig:pot_beta3}
\begin{center}
\end{center}
\end{figure*}

In the case of $\beta _{3}\neq 0$ analytical solutions to Eq.~(\ref{ODE})
are not available. The key goal of the numerical solution of this equation
was to produce the \textquotedblleft most irregular" (nonperiodic) evolution
of $|u(t)|^{2}$ that remains bounded (does not develop singularities).

Before proceeding to numerical results, we start with an analysis of initial
conditions in the case of $\beta _{3}=0$, i.e. when Eq.~(\ref{ODE}) is
integrable. In terms of variables $x\equiv a_{1}$, $y\equiv a_{2}$ and $%
p_{x}\equiv b_{1}$, $p_{y}\equiv b_{2}$, Hamiltonian (\ref{H}) with $\beta
_{3}=0$ is
\begin{equation}
H=-\left[ \beta _{2}\frac{p_{x}^{2}+p_{y}^{2}}{2}+k(x^{2}+y^{2})-\frac{%
(x^{2}+y^{2})^{2}}{2}\right] ,  \label{newhxy}
\end{equation}%
which is tantamount to the Hamiltonian of a particle moving in the 2D plane $%
\left( x,y\right) $ under the action of a rotationally symmetric potential.
The angular momentum of such a particle, $L=xp_{y}-yp_{x}$ is conserved ($%
L\equiv -I/\beta _{2}$, where $I$ is dynamical invariant (\ref{I}) in the
case of $\beta _{3}=0$). Consequently, the time evolution of $%
|u(t)|^{2}\equiv x^{2}(t)+y^{2}(t)$, which determines the external potential
for the signal channel, can be either periodic or unbounded (as mentioned
above), unbounded solutions being possible for $\beta _{2}>0$. In what
follows below, while numerically integrating Eq.~(\ref{ODE}) with $\beta
_{2}>0$, we chose initial conditions so that the motion was bounded for $%
\beta _{3}=0$.

It might seem, formally, that choosing the anomalous-GVD sign, $\beta _{2}<0$%
, in the support channel, which gives rise to the classical bright solitons
\cite{ZS,Agrawal} and ensures the boundedness of solutions for $\beta _{3}=0$%
, might be the best choice, but this is not the case. When $\beta _{3}\neq 0$%
, our numerical results demonstrate that the maximum value of $\beta _{3}$
that secures the absence of singularities in the solutions is slightly
larger for $\beta _{2}>0$ (the normal-GVD sign) than for $\beta _{2}<0$ (at
least, in the numerical searches that we have conducted). Hence, the choice
of $\beta _{2}>0$ is more appropriate for producing nonsingular QP
solutions. Another important argument in favor of the choice of the normal
GVD is that (quasi-) periodic solutions in the case of anomalous GVD are
subject to the modulational instability, which tends to split the
(quasiperiodic) wave into a chain of bright solitons \cite{MI1,MI2}, thus
making the scheme inappropriate for the experimental realization, while this
instability is typically absent in the case of the normal GVD.

Figure~\ref{fig:pot_beta3} shows typical examples of solutions for the
optical power in the support channel, $|u(t)|^{2}$, along with their Fourier
transforms, which are defined in the usual form,%
\begin{equation}
F(\omega )=\int_{-\infty }^{+\infty }e^{-i\omega t}|u(t)|^{2}dt,
\label{Fourier}
\end{equation}%
(actually, the integration in Eq.~(\ref{Fourier}) is performed over the
temporal range in which the numerical solution was produced). 
For $\beta_3 \ne 0$, the system is not integrable, and the evolution of $|u(t)|^2$ is not periodic for generic initial conditions. However, in the cases of $\beta_3 = 0.05$ and $0.1$, presented in the left and central columns of Fig.~\ref{fig:pot_beta3}, the quasi-periodic behavior is very weak. This is confirmed by the presence of strong peaks only at the main frequency and its higher harmonics in the spectra plots (note that the spectra are presented on a logarithmic scale). For $\beta_3 = 0.15$, shown in the right column of the figure, a well-pronounced quasi-periodic pattern emerges. This pattern is used throughout the rest of the paper to clearly demonstrate the AA localization in the signal channel.

It is relevant to stress that the numerical solution of Eq.~(\ref{ODE})
corroborates the conservation of the dynamical invariants (\ref{I}) and (\ref%
{H}), as predicted by the above analysis.


\subsection{The Aubry-Andr\'{e}-like localization of the signal wave in the
temporal domain}

Here we move on to the core part of the work, which demonstrates the
temporal localization of the signal field $\Psi $ governed by Eq.~(\ref{Psi}%
) under the action of the QP supporting structure $|U(z,t)|^{2}=|u(t)|^{2}$
analyzed in the previous section. To this end, we look for solutions to Eq. (%
\ref{Psi}) with a real propagation constant $E$,
\begin{equation}
\Psi (z,t)=e^{iEz}\psi (t)  \label{Psipsi}
\end{equation}%
(cf. Eq. (\ref{UV})), reducing the propagation equation (\ref{Psi}) to the
ODE,
\begin{equation}
\left[ -\frac{\gamma _{2}}{2}\frac{\partial ^{2}}{\partial t^{2}}-\frac{%
i\gamma _{3}}{6}\frac{\partial ^{3}}{\partial t^{3}}+2|u(t)|^{2}\right] \psi
=E\psi .  \label{eqPsi1}
\end{equation}%
This equation with $\gamma _{3}=0$ is tantamount to the standard 1D linear
Schr\"{o}dinger equation with the temporal coordinate $t$, effective mass $%
1/\gamma _{2}$ and QP potential $2|u(t)|^{2}$, that can give rise to the
AA-like localization. In particular, comparing Eq.~(\ref{eqPsi1}) to the
discrete (tight-binding) AA model, we conclude that parameter $\gamma _{2}$
in Eq.~(\ref{eqPsi1}) is proportional to the inter-site tunneling rate in
the AA model. As concerns the third-order GVD, represented by $\beta _{3}>0$%
, it is crucially important for the production of the QP solutions by Eq.~(%
\ref{ODE}), but the similar term $\sim \gamma _{3}$ in Eq.~(\ref{eqPsi1}) is
not a major factor for the study of the localization effect in the framework
of the latter equation, therefore it will be considered elsewhere.

Proceeding to the detailed analysis, we first consider the spectrum of
eigenvalues $E$ produced by numerical solution of Eq. (\ref{eqPsi1}) with
the open (Dirichlet) boundary conditions, i.e., $\psi (t=0)=\psi (t=T)=0$,
where $T$ is much larger than the time scale over which $|u(t)|^{2}$ varies.
In this context, we discretize the time coordinate as $t_{n}=ndt$, with
integer $n$ and sufficiently small $dt$, to achieve well-converging results,
and represent the left-hand side of Eq.~(\ref{eqPsi1}) in the form of an
Hermitian matrix,
\begin{eqnarray}
-\frac{\gamma _{2}}{2dt^2}\left[\psi(t_{n+1})-2\psi(t_n)+\psi(t_{n-1})\right]
\cr 
+2|u(t_n)|^{2} \psi(t_n)
=E\psi(t_n),  \label{eqPsi1_discrete}
\end{eqnarray}%
which can be diagonalized by standard procedures.

\begin{figure}[t]
\centering
\includegraphics[width=0.48\textwidth]{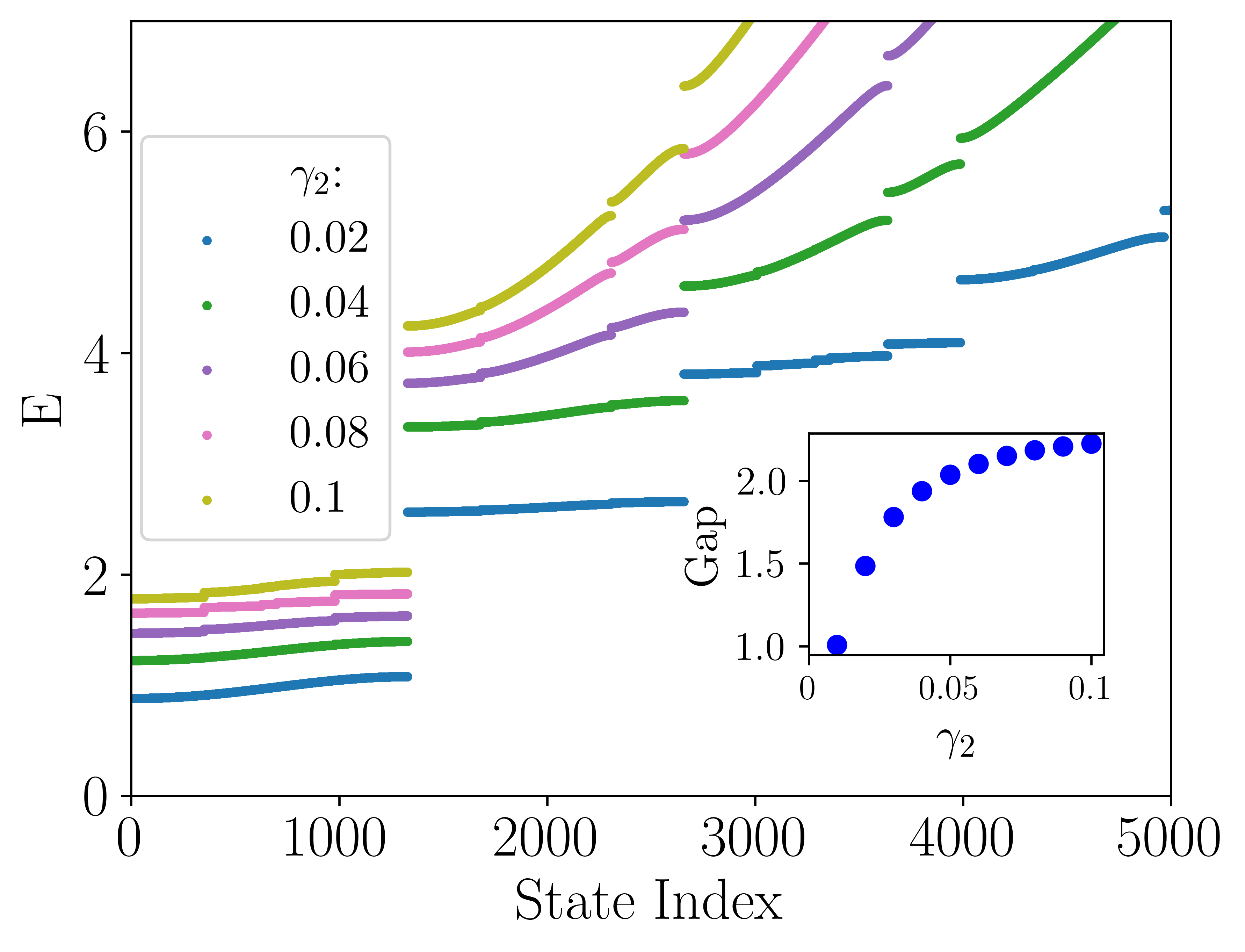}
\caption{Eigenvalues $E$, in the ascending order, produced by the numerical
solution of Eq.~(\protect\ref{eqPsi1}) with $\protect\gamma _{3}=0$ (the
same is set in all figures produced below) and the effective potential $%
2|u(t)|^{2}$, which is shown in the right column of Fig.~\protect\ref%
{fig:pot_beta3}. The solutions were obtained with the open boundary
conditions, i.e. $\protect\psi (=0)=\protect\psi (t=T\equiv 1000)=0$. Values
of $\protect\gamma _{2}$ are indicated in the panels. In all cases, the band
structure is evident. The inset shows the gap between the first and second
bands as a function of. $\protect\gamma _{2}$.}
\label{fig:band_structure}
\end{figure}

Figure~\ref{fig:band_structure} shows that the eigenvalues form a band
structure with gaps that become larger as $\gamma _{2}$ increases. Examples
of eigenstates from the first and second bands are shown in Fig.~\ref%
{fig:loc_fit}. For sufficiently small values of $\gamma _{2}$ (i.e.,
relatively large values of the effective mass in Eq.~(\ref{eqPsi1})), the
QP-shaped effective potential, $2|u(t)|^{2}$, indeed produces exponentially
confined eigenstates $\psi (t)$, which precisely exhibit the AA localization
in the temporal domain. By positioning an optical detector at some fixed
location $z$, one will observe that the temporarily localized signal $\psi
(t)$ gradually emerges from zero and then exponentially decays at large
times.

\begin{figure*}[t]
\centering
\includegraphics[width=0.341\textwidth]{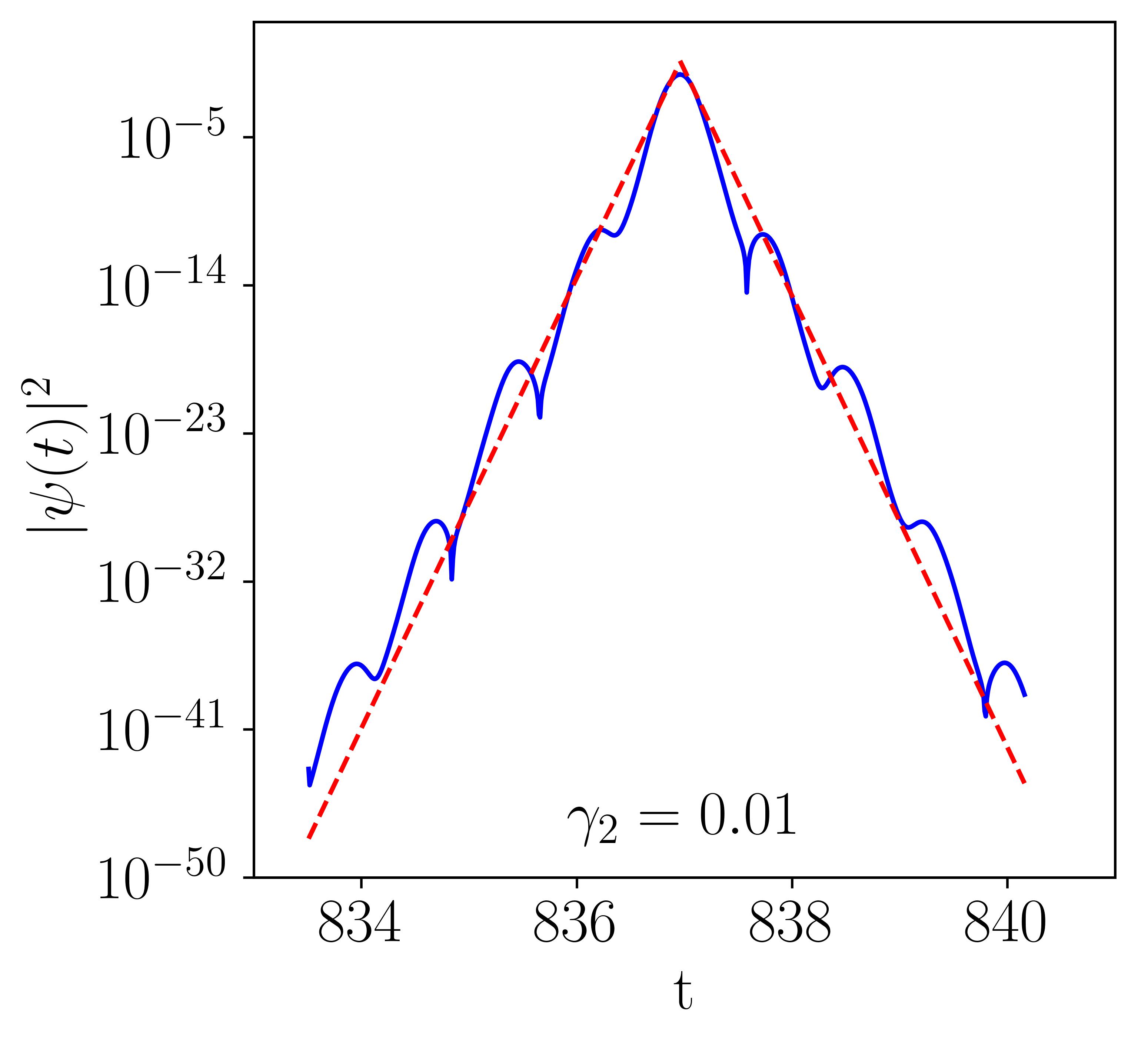} %
\includegraphics[width=0.315\textwidth]{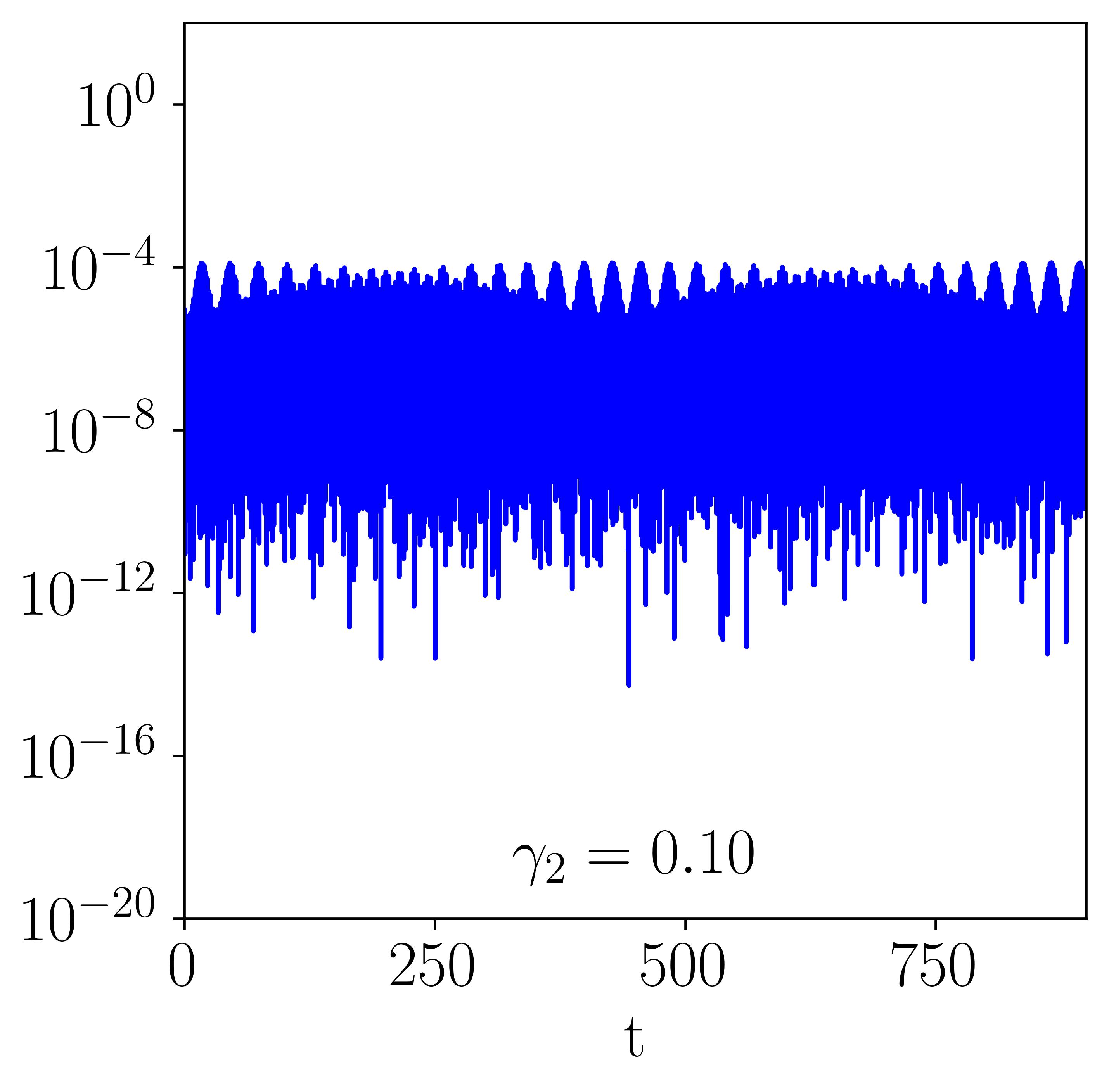} %
\includegraphics[width=0.32\textwidth]{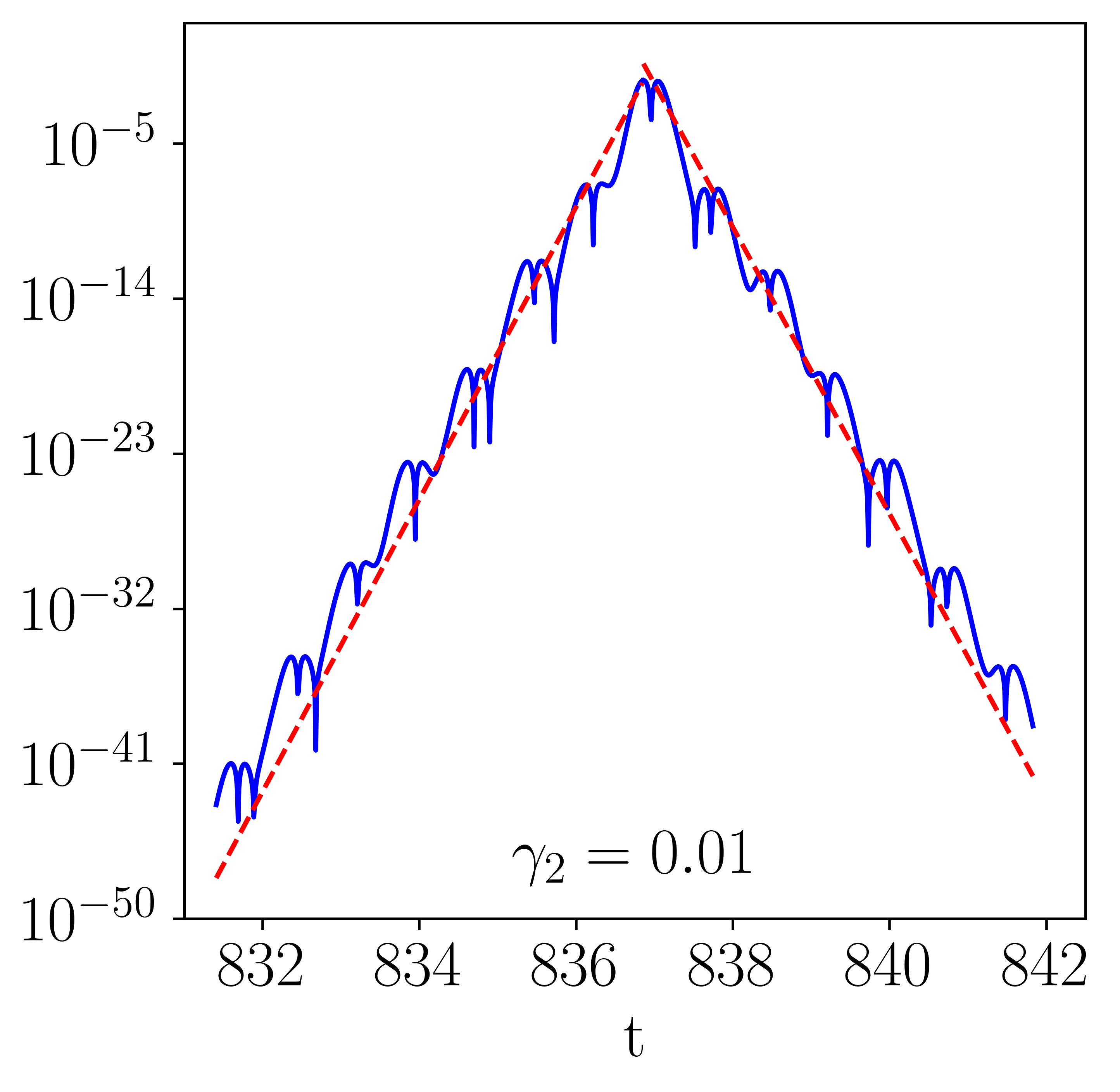}
\caption{The left panel shows $\left\vert \protect\psi (t)\right\vert ^{2}$,
on the logarithmic scale, for the well-localized eigenstate of Eq.~(\protect
\ref{eqPsi1}) corresponding to eigenvalue $E$ in the middle of the first
band for $\protect\gamma _{2}=0.01$, see Fig.~\protect\ref%
{fig:band_structure}. Similarly, the middle panel shows the eigenstate for $%
\protect\gamma _{2}=0.1$, where no signature of the localization is
observed. The right panel presents the eigenstate corresponding to
eigenvalue $E$ in the middle of the second band for $\protect\gamma %
_{2}=0.01 $ (the same as in the left panel). For the same $\protect\gamma %
_{2}=0.1$ as in the middle panel, the respective eigenstate shows no
localization signature either in the second band. In the left and right
panels, we also plot exponential profiles $B\exp \left( -|t-t_{0}|/\protect%
\xi \right) $ with parameters fitted to the plots of the localized
eigenstates. The so obtained values of parameter $\protect\xi $ agree very
well with the localization time calculated with the help of the\
transfer-matrix approach. Namely, the fitted values are $\protect\xi =0.0752$
and $0.1183$ for the left and right panels, while the corresponding
transfer-matrix results are $0.0727$ and $0.1175$, respectively. All other
parameters are the same as in Fig.~\protect\ref{fig:band_structure}.}
\label{fig:loc_fit}
\end{figure*}

The localized eigenstates produced by the numerical solution of Eq.~(\ref%
{eqPsi1}) decay exponentially. This feature is clearly corroborated, in the
left and right panels of Fig.~\ref{fig:loc_fit}, by the fitted exponential
profiles, $B\exp \left( -|t-t_{0}|/\xi \right) $. The localization time $\xi
$ (the counterpart of the localization length in the spatial domain) can be
determined using the transfer-matrix approach \cite{Derrida1984}. To this
end, we use the discrete representation of functions, $\psi _{n}=\psi (ndt)$%
, similarly to the numerical diagonalization of Eq.~(\ref{eqPsi1}). Then,
using Eq.~(\ref{eqPsi1}), we compute
\begin{equation}
R_{n}\equiv \frac{\psi _{n}}{\psi _{n-1}}=\frac{2\left( dt\right) ^{2}}{%
\gamma _{2}}(2|u_{n}|^{2}-E)+2-\frac{1}{R_{n-1}},  \label{R}
\end{equation}%
where the discretized support structure is $|u_{n}|^{2}=|u(ndt)|^{2}$, and
we iterate $R_{n}$ as per Eq.~(\ref{R}),\ starting from $R_{1}\neq 0$. The
localization time is thus obtained as $\xi =\lim_{N\rightarrow \infty }\left[
Ndt\left( \sum_{n=1}^{N}\log |R_{n}|\right) ^{-1}\right] $ \cite{Derrida1984}%
. The numerically calculated localization times closely match the values
obtained from the fitting of the exponential profiles, see Fig.~\ref%
{fig:loc_fit}.

\begin{figure}[t]
\centering
\includegraphics[width=0.48\textwidth]{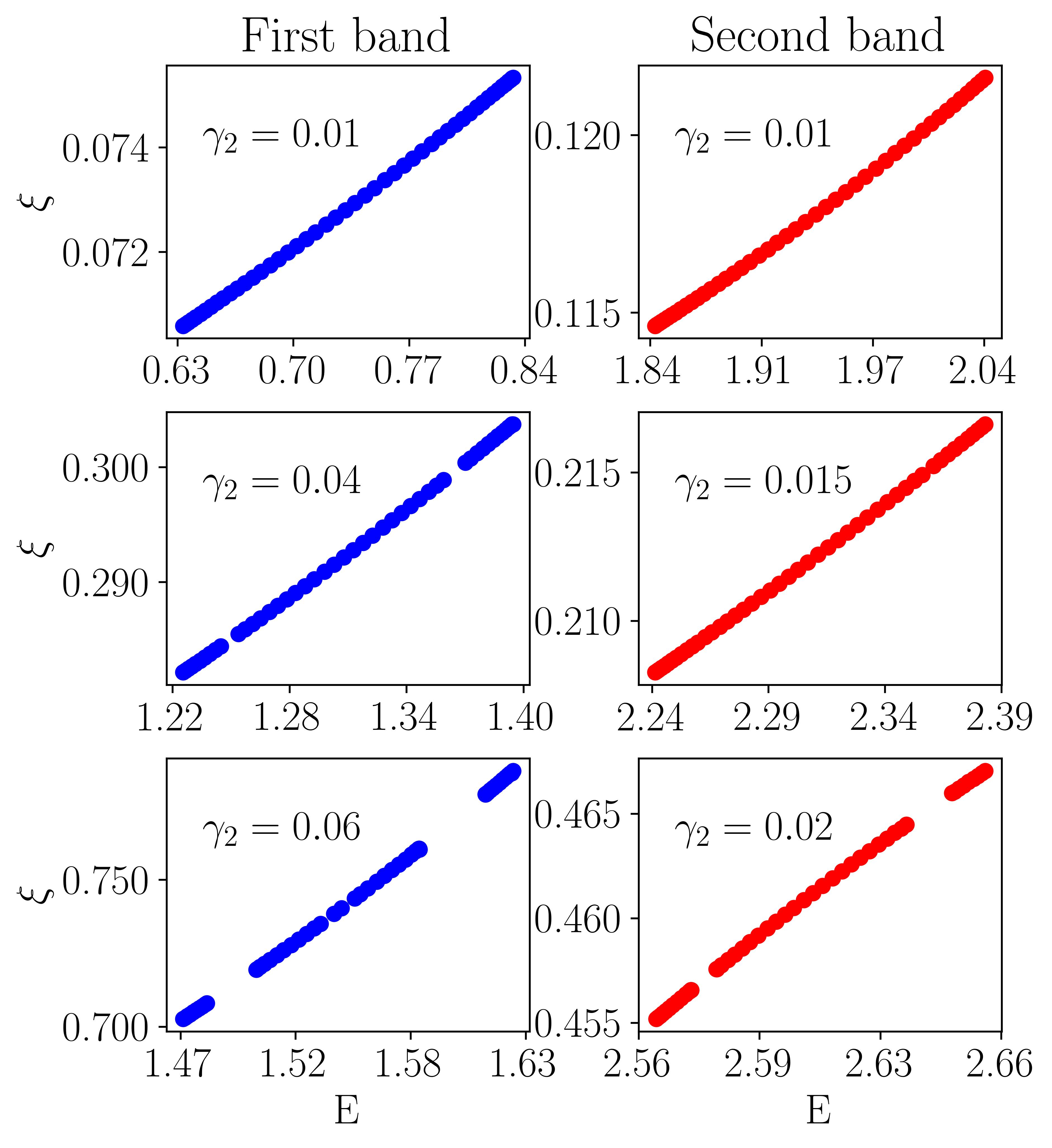}
\caption{Localization time $\protect\xi $ vs. eigenvalue $E$, as produced by
Eq.~(\protect\ref{eqPsi1}), using the transfer-matrix calculations (see Eq.~(%
\protect\ref{R}))\ in the first and second bands (the left and right
columns, respectively), for different values of $\protect\gamma _{2}$, as
indicated in the panels.  For certain ranges of $E$, small gaps in the eigenvalues are present within the bands (cf. Fig.~\ref{fig:band_structure}). In these gaps, the localization time is not provided (no points on the plots) because there are no solutions corresponding to these eigenvalue ranges. Other parameters are the same as in Fig.~\protect
\ref{fig:band_structure}.}
\label{fig:loc_state}
\end{figure}

\begin{figure}[t]
\centering
\includegraphics[width=0.48\textwidth]{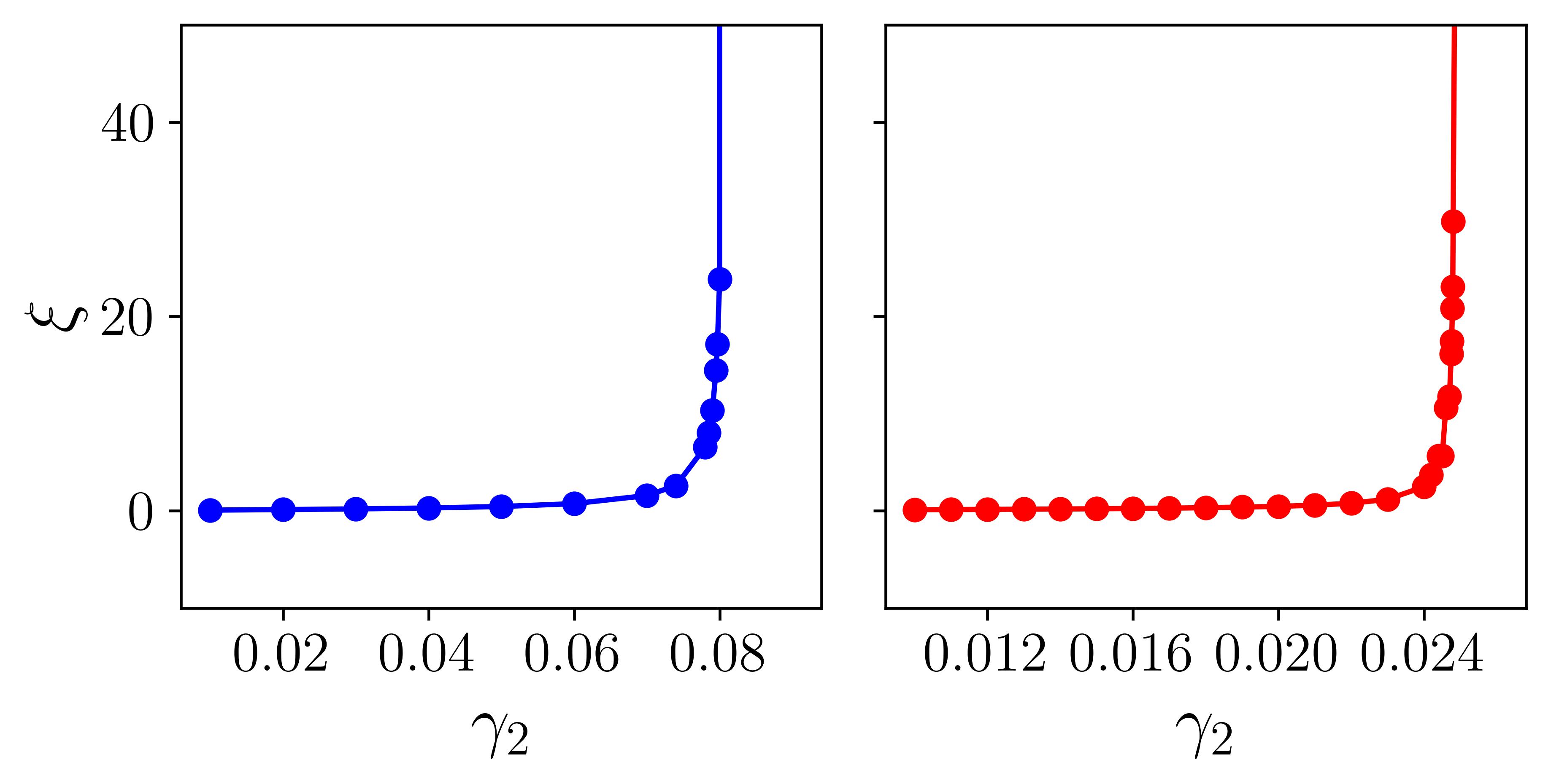}
\caption{Localization time $\protect\xi $ vs. the GVD coefficient $\protect%
\gamma _{2}$ in the first and second bands (the left and right panels,
respectively). States from the middle of the bands were chosen but similar
behavior is observed in the entire bands. All other parameters are the same
as in Fig.~\protect\ref{fig:band_structure}.}
\label{fig:loc_state1}
\end{figure}

To summarize the results produced by the systematic analysis of the present
system, Fig.~\ref{fig:loc_state} shows the localization time $\xi $,
identified by means of the transfer-matrix method in the first and second
bands at different values of $\gamma _{2}$. It is seen that, in both bands, $%
\xi $ changes slightly, being weakly sensitive to the variation of
eigenvalue $E$. On the other hand, Fig.~\ref{fig:loc_state1} exhibits a
steep change of $\xi $ following the variation of the GVD coefficient $%
\gamma _{2}$. The figure reveals that, in both bands, there is a critical
value of $\gamma _{2}$, above which the numerically obtained localization
time $\xi $ diverges, i.e., the localization phenomenon vanishes. This
happens when the trend to spreading (delocalization) of the wave function $%
\psi (t)$ becomes too strong with the growth of coefficient $\gamma _{2}$ in
Eq.~(\ref{eqPsi1}). Such critical behavior is qualitatively similar to that
in the AA model.

\section{Conclusion}

\label{sec4}

The aim of this work is to propose a new physical setup for the realization
of a particular manifestation of the time-crystal behavior, \textit{viz}.,
temporal localization of the Aubry-Andr\'{e} type of linear signals in
optical fibers under the action of an effective QP (quasiperiodic)
potential, which is induced through the XPM (cross-phase-modulation) effect
by a nonlinear QP wave copropagating \ in the support channel in the same
fiber. The QP waves are possible when the third-order GVD (group-velocity
dispersion) is taken into regard in the support channel. In particular, two
dynamical invariants of the six-dimensional dynamical system governing the
wave propagation in the support channel are found in the exact form. The QP
patterns in the support channel and solutions for the temporarily localized
optical pulses in the signal channel are produced in the numerical form.
Properties of the localized pulses, and the boundary of their delocalization
are investigated in detail.

The possibility of the experimental realization of the scheme in fiber optics is supported by essentially the same analysis of the physical setting which was presented in terms of the two-channel system in Refs. \cite{Ship1} and \cite{Ship2}. The temporal localization of the signal pulses reported in the present work may find an application to fiber-optic communications, as it offers a possibility to maintain the \textit{return-to-zero} regime of the data-stream transmission \cite{Agrawal} in the low-power linear regime.

As an extension of the work, it may be interesting to consider effects of
nonlinearity and third-order GVD in the signal channel. It is expected that,
depending on the sign of GVD in this channel, the nonlinearity may enhance
the localization (leading to the formation of bright solitons \cite{Agrawal}%
) or stimulate delocalization \cite{Flach}.

\section*{Acknowledgments}

This research was funded by the National Science Centre, Poland, Project No.
2021/42/A/ST2/00017. The work of B.A.M. was supported, in part, by the
Israel Science Foundation through grant No. 1695.22.



%

\end{document}